\documentclass{mem}
\usepackage{natbib}\usepackage{txfonts}\usepackage{balance}
\usepackage{graphicx}
\usepackage[a4paper]{hyperref}
\idline{75}{282}
\begin{document}
\def\teff{$T\rm_{eff }$}
\def\kms{$\mathrm {km s}^{-1}$}

\title{
True solar analogues in the open cluster M67
}

   \subtitle{}

\author{
K. \,Biazzo\inst{1,2} 
\and L. \, Pasquini\inst{2}
\and P. \, Bonifacio\inst{3,4,5}
\and S. \, Randich\inst{6}
\and L. R. \, Bedin\inst{7}
          }

  \offprints{K. Biazzo \email{katia.biazzo@oact.inaf.it}}

\institute{
INAF - Catania Astrophysical Observatory, Catania, Italy
\and
ESO - European Southern Observatory, Garching bei M\"unchen, Germany
\and
CIFIST, Marie Curie Excellence Team
\and
GEPI, Observatoire de Paris, Meudon, France
\and
INAF - Trieste Astronomical Observatory, Trieste, Italy
\and
INAF - Arcetri Astrophysical Observatory, Arcetri, Italy
\and
Space Telescope Science Institute, Baltimore, United States
}

\authorrunning{K. Biazzo et al.}

\titlerunning{True solar analogues in M67}
\abstract{The solar analogues are fundamental targets for a better understanding of our
Sun and Solar System. Notwithstanding the efforts, this research
is usually limited to field stars. The open cluster M67 offers a unique
opportunity to search for solar analogues because its chemical composition and
age are very similar to those of our star. In this work, we analyze FLAMES@VLT spectra of about one hundred
of M67 main sequence stars with the aim to identify solar analogues. 
We first determine cluster members which are likely not
binaries, by combining both proper motions and radial velocity measurements.
Then, we concentrate our analysis on the determination of stellar effective
temperature, using the analyzes of line-depth ratios and H$\alpha$ wings.
Finally, we also compute lithium abundance for all the stars. 
Thanks to the our analysis, we find ten solar analogues, which allow us
to derive a solar color $(B-V)=0.649\pm0.016$ and a cluster distance modulus of $9.63\pm0.08$,
very close to values found by previous authors. Among them, five are the best
solar twins with temperature determinations within 60 K from the solar values. 
Our results lead us to do further spectroscopic investigations because the solar  
analogues candidates are suitable for planet search.

\keywords{Stars: fundamental parameters -- Open clusters and associations: individual: M67 
-- Stars: late-type}}
\maketitle{}

\section{Introduction}
The search for exo-planets has been mostly developed around field stars. Such stars 
present several advantages, for instance a wide range of stellar characteristics 
(mass, age, effective temperature, chemical composition, etc.), which allows us to 
study the dependence of planet formation on stellar parameters. 

Another line of research is the specificity 
of our Sun and the opportunity to find solar stars hosting exo-planets. This search 
is best performed in open clusters, which, showing homogeneous age and chemical 
composition, common birth and early dynamical environment (\citealt{Randich2005}), 
provide us an excellent laboratory for investigating the physics of planetary system 
formation. The old open cluster M67 is to this purpose a perfect target, having many 
solar-type stars and showing an age encompassing that of the Sun ($3.5-4.8$ Gyrs; 
\citealt{Yadav2008}), 
a solar metallicity ([Fe/H]=$0.03\pm0.02$, \citealt{Randich2006}) and lithium depleted 
G stars (\citealt{Pasquini1997}). 

The present paper is the culmination  of a work, which involved the chemical determination of this cluster 
(\citealt{Randich2006}), photometry and astrometry (\citealt{Yadav2008}) to obtain membership, and 
FLAMES/GIRAFFE high resolution spectroscopy to clean this sample from binaries, and to look for the best solar 
analogues using the line-depth ratios method and the wings of the H$\alpha$ line to 
determine accurate temperatures with respect to the Sun (\citealt{Pasquini2008}).

\section{Observations} 
We observed M67 for 2.5 hours during three nights in February 2007 with the 
multi-object FLAMES/GIRAFFE spectrograph at the UT2/Kueyen ESO-VLT in Paranal 
(Chile). We chose the HR15N MEDUSA mode, which allows us to cover the spectral 
range 6470-6790 \AA~and, consequently, to cover the H$\alpha$ and the lithium lines. 
With this configuration, the resolution of 17\,000 gives us the possibility to obtain 
for almost 100 stars good radial velocities and to perform effective temperature and 
lithium abundance measurements. 

We have chosen from the catalogue of \cite{Yadav2008} bright stars ($13\fm0<V<15\fm0$) 
with $(B-V)$ in the solar neighbor ($0.60-0.75$) 
which shown the best combination of proper motions measurements 
($\mu_{\alpha}\cos \delta$, $\mu_{\delta}$) and 
proper-motion membership probability ($P_{\mu}$) allowing us to observe at a time almost 100 
stars to be observed with FLAMES/GIRAFFE (\citealt{Pasquini2008}). 

\section{Results}
\subsection{Radial Velocity}
From the radial velocity variations of the 90 stars observed in three nights with FLAMES/GIRAFFE, 
we find that 59 of them are probable single cluster members with an average radial velocity 
$<V_{\rm rad}>$= 32.9 km s$^{-1}$ and a $\sigma$=0.73 km s$^{-1}$ (\citealt{Pasquini2008}). 

\subsection{Lithium Abundance}
Since the lithium element is likely a `thermometer' of the complex interaction taking place in the 
past between the stellar external layers and the hotter interior, we have computed the equivalent 
width of the lithium line at $\lambda$=6707.8 \AA. One of the important points characterizing M67 
is that we find many main sequence stars sharing the same lithium abundance of the Sun, indicating 
a similar mixing history. Moreover, for the first time, we see the Li extra-depletion appears 
in stars cooler than 6000 K (\citealt{Pasquini2008}).

\subsection{Effective Temperature}
\begin{figure*}[ht]
\center
\includegraphics[width=6.7cm]{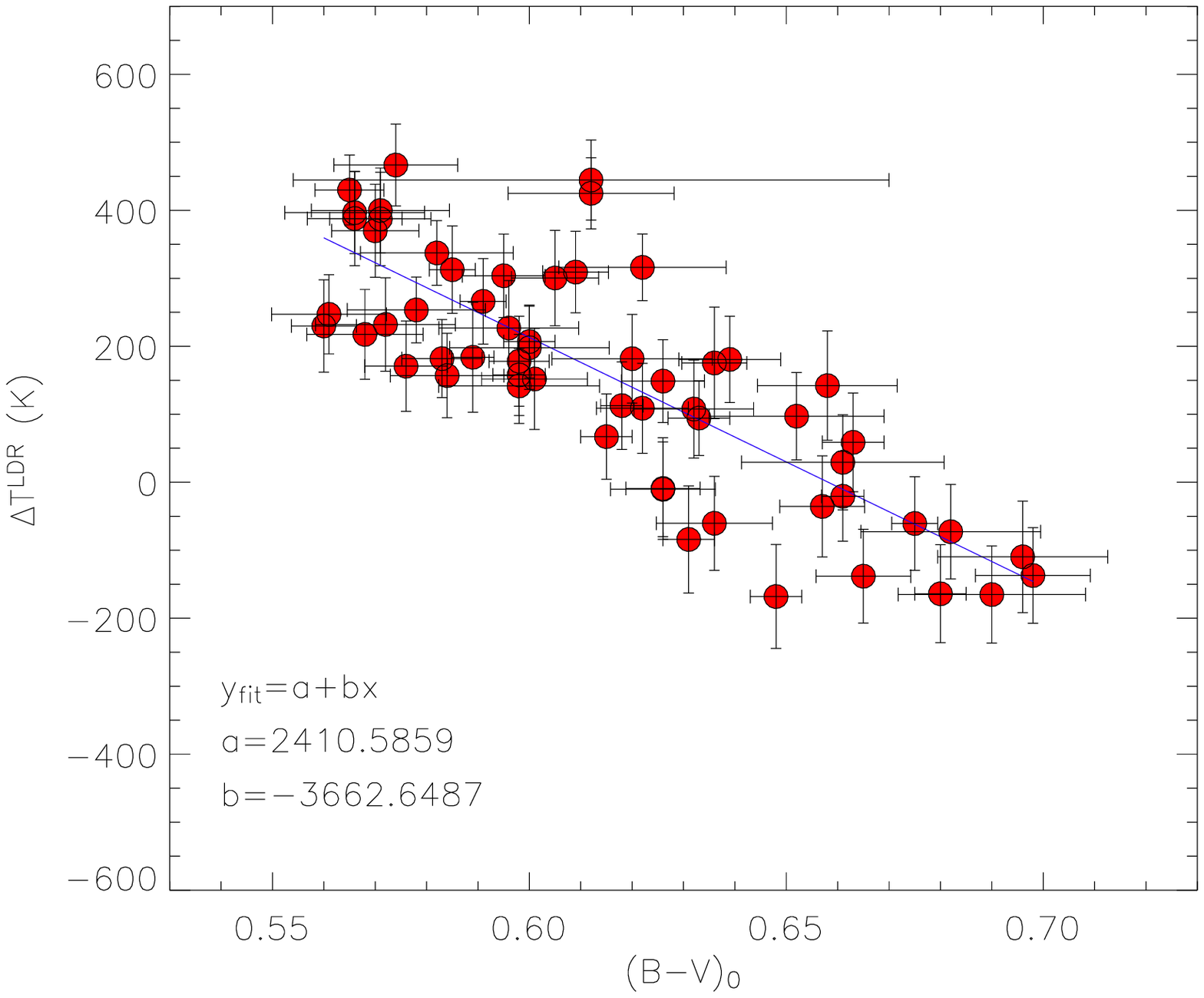}
\includegraphics[width=6.7cm]{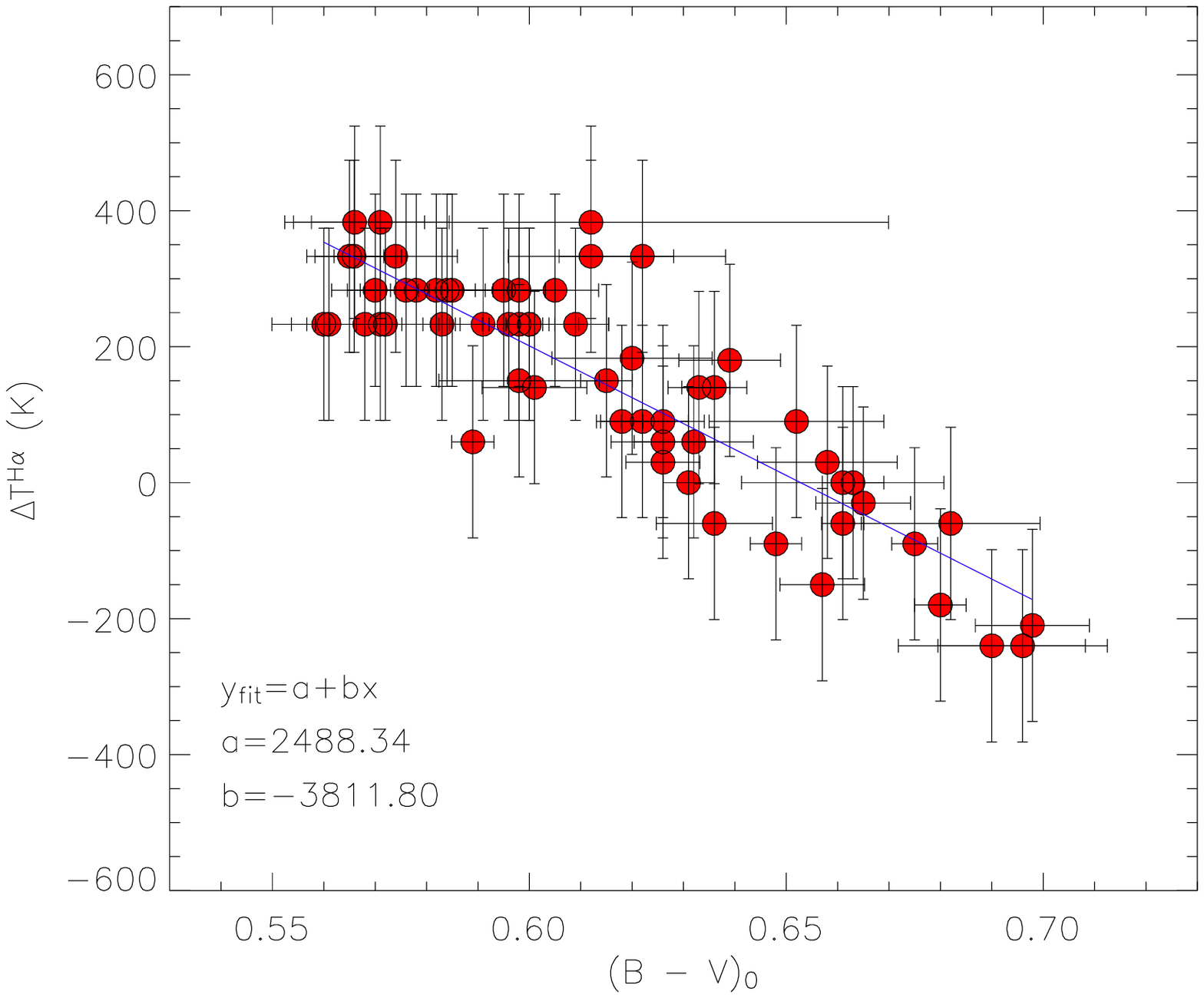}
\caption{\footnotesize {\it Left panel:} Results of the analysis of six line pairs and the three observing 
runs. {\it Right panel:} Results of the analysis of H$\alpha$ wings on the summed GIRAFFE spectra. 
The two graphs represent the difference in temperature between the stars and the Sun, as a function of 
the de-reddened $(B-V)$.}
\label{fig:deltaT}
\end{figure*}

Thanks to synthetic spectra computed in the temperature range between $5400-6300$ K, we have analyzed, 
in the spectral region covered by FLAMES/GIRAFFE, which lines were sensitive to temperature. At the end, 
we have selected six couples of them and applied a method based on line-depth ratios (LDRs) to derive 
the effective temperature of the probable members (\citealt{GrayJoha1991,Catalano2002,Biazzo2007a,Biazzo2007b}). 
Thus, we have developed appropriate LDR$-T_{\rm eff}$ calibrations on synthetic spectra 
and derived the effective temperature of the probable members. 
Fig.~\ref{fig:deltaT} shows the temperature difference $\Delta T^{\rm LDR}$ between the 
FLAMES/GIRAFFE targets and the Sun obtained from LDR method as a function of the 
de-reddened $(B-V)$ color.

Since the wings of the H$\alpha$ line profile are very sensitive to temperature, we have also studied the 
behavior of this diagnostics. According to \cite{Cayrel1985}, the effective temperature 
of a star can be derived from the strength of its H$\alpha$ wings measured between 3 and 5 \AA~from 
the H$\alpha$ line-center, as compared to synthetic spectra H$\alpha$ line-wings in the same wavelength 
interval. Figure~\ref{fig:deltaT} shows the relationship between the temperature difference 
$\Delta T^{\rm H\alpha}$ of the FLAMES/GIRAFFE targets and the Sun as obtained from the H$\alpha$ wings, 
and the de-reddened $(B-V)$ color.

\section{Solar analogues}
\begin{figure*}	
\center
\includegraphics[width=8.5cm]{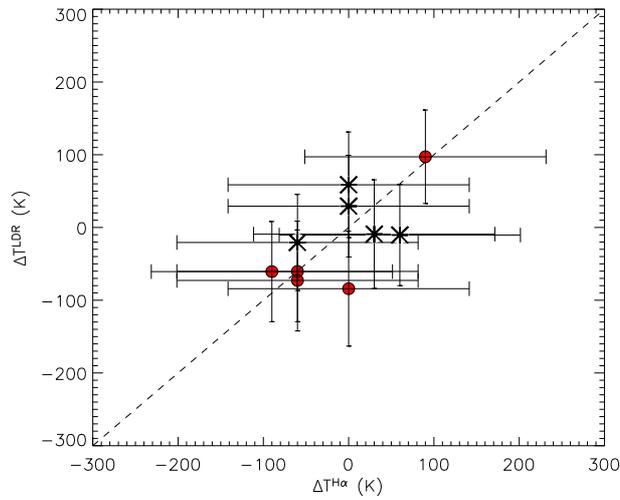}
\caption{\footnotesize Difference in temperature between the 10 best solar analogues and the Sun. 
The ordinate is referred to the temperature derived thanks to the LDR technique, while the abscissa 
is referred to the temperature derived thanks to the method based on the H$\alpha$ wings. The asterisks 
represent the positions of the five best solar twins (\citealt{Pasquini2008}).
}
\label{fig:deltaT_prop}
\end{figure*}

Comparing the $\Delta T^{\rm LDR}$ and the $\Delta T^{\rm H\alpha}$ and taking into account our lithium 
abundance determination, we find ten solar analogues. In particular, five stars are the 
closest to the Sun, with effective temperature derived with both methods within 60 K from the solar one 
(Fig.~\ref{fig:deltaT_prop}). 

\begin{figure*}	
\center
\includegraphics[width=10cm]{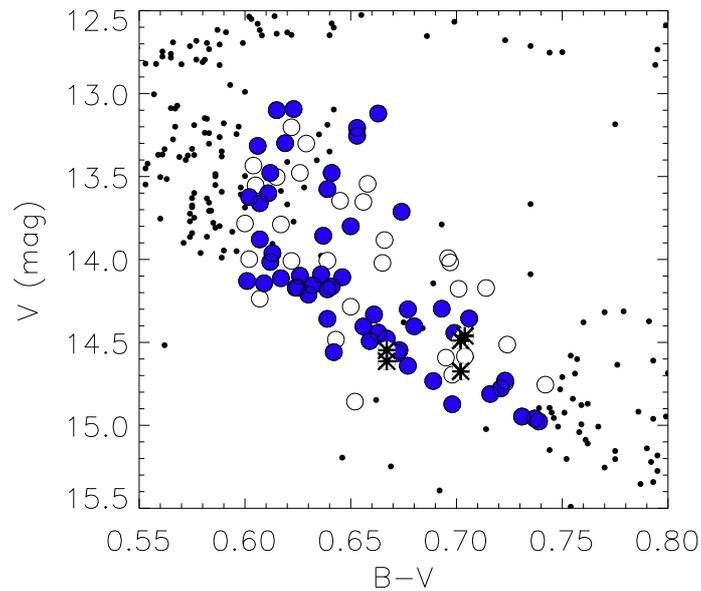}
\caption{\footnotesize Zoom of the M67 color-magnitude diagram, centered on the 90 targets observed. 
With empty circles the 59 retained single member candidates, while with filled circles the stars discarded 
are shown (figure taken from \citealt{Pasquini2008}). The 5 best solar twins are marked with asterisks.
}
\label{fig:zoom}
\end{figure*}

Figure~\ref{fig:zoom} shows 
a zoom of the M67 color-magnitude diagram centered on the all sample observed. The 
59 single members and the 5 best solar twins are marked with different colors (\citealt{Pasquini2008}).

Moreover, through a inversion technique, our solar analogues allowed us to obtain a precise estimate 
of the solar $(B-V)$ and accurate cluster distance modulus. Our final results are $(B-V)_{\odot}=0.649\pm0.016$ 
and $V-M_{\rm V}=9.63\pm0.08$, which are in excellent agreement with the most recent determinations. 

\section{Conclusions}
By using spectroscopic observations performed with FLAMES/GIRAFFE at the VLT, we have found the best 
solar twins in M67 thanks to accurate determinations of radial velocity, lithium abundance, and 
effective temperature. 

Thanks to our promising results, we plan to apply our method to other open clusters younger and 
older than the Sun in order to get informations about ``progenitors'' and ``descendants'' of our unique star.

Our method has proved to be suitable for the determination of both the solar color and the cluster 
distance modulus. It could therefore be applied for the same purpose to other open clusters.

\bibliographystyle{aa}

\end{document}